\documentclass[a4paper]{jpconf}
\usepackage{graphicx}

\newcommand{\cuga}{$ \mathrm{CuGa_{2}O_{4}} $}
\newcommand{\cual}{$ \mathrm{CuAl_{2}O_{4}} $}

\begin{document}
\title{Zero-point entropy of the spinel spin glasses CuGa$_2$O$_4$ and CuAl$_2$O$_4$}

\author{L A Fenner$^1$, A S Wills$^{1,2}$, S T Bramwell$^2$, M Dahlberg$^3$ and P Schiffer$^3$}

\address{$^1$ Chemistry Department, UCL, 20 Gordon Street, London WC1H 0AJ, UK}
\address{$^2$ London Centre for Nanotechnology, 17-19 Gordon Street, London WC1H 0AH, UK}
\address{$^3$ Department of Physics and Materials Research Institute, Pennsylvania State University, University Park, Pennsylvania 16802, USA}

\ead{a.s.wills@ucl.ac.uk}

\begin{abstract}
The zero-point entropy of a spin glass is a difficult property to experimentally determine and interpret. Spin glass theory provides various predictions, including unphysical ones, for the value of the zero-point entropy, however experimental results have been lacking. We have investigated the magnetic properties and zero-point entropy of two spinel $\mathrm{Cu^{2+}}$ based spin glasses, \cuga\ and \cual. Dc- and ac-susceptibility and specific heat measurements show many characteristic spin glass features for both materials. The spin glass freezing temperature is determined to be $T_{\mathrm{f}} = 2.89 \pm 0.05\,\mathrm{K}$ for \cuga\ and $T_{\mathrm{f}} = 2.30 \pm 0.05\,\mathrm{K}$ for \cual. By integrating the specific heat data we have found that \cuga\ and \cual\ have zero-point entropies of $S_{0} = \mathrm{4.96\,JK^{-1}mol^{-1}}$ and $S_{0} = \mathrm{4.76\,JK^{-1}mol^{-1}}$ respectively. These values are closest to the prediction for a Sherrington-Kirkpatrick XY spin glass, however they are notably higher than all of the theoretical predictions. This indicates that \cuga\ and \cual\ have a greater degeneracy in their ground states than any of the spin glass models.
\end{abstract}

\section{Introduction}
Nernst's theorum states that the entropy of a material at absolute zero is zero. However, this condition does not apply to systems which are not in true thermodynamic equilibrium, {\it e.g.} glasses \cite{BRA06}. The energy of such systems depends not only on temperature, but also on the sample history, and this can result in there being residual entropy at 0\,K \cite{LAN90}. This zero-point entropy is a key feature of frustrated systems, as it characterises the ground state complexity.

Normal water ice \cite{PAU35,GIA36}, and the pyrochlore \cite{HAR97,RAM99} and kagome \cite{kag_spin_ice,Cumings} spin ices are all example systems where the experimental value of the zero-point entropy has been found to agree well with prediction.  Spin glasses are another family of frustrated magnetic systems which are believed to possess macroscopically degenerate ground states. The transition from a paramagnetic state to a spin glass state at a temperature $T_\mathrm{f}$ involves the gradual freezing of the fluctuating magnetic moments into a random arrangement.  Experimentally, both conventional spin glasses based on random exchange or site disorder \cite{MYD93}, and exotic `topological' \cite{Ritchey:1993,H3O_Fe} spin glasses have been found. Despite this, there has been a lack of experimental evidence to confirm or disprove predictions of the zero-point entropy, partly due to the intrinsic complexity of spin glasses. For example, they do not have the self-averaging property shown by spin ice, making it more difficult to correctly interpret and understand any residual entropy data.

In this work we have studied the spin glass properties and residual entropy of two insulating, spinel spin glasses, \cuga\ and \cual, and have found that the residual entropies do not agree with any of the proposed values.

\section{Experimental Details}
Polycrystalline samples of \cuga\ and \cual\ were prepared by a solid state technique from appropriate amounts of the parent oxides. Reactions, carried out at 950$^{\circ}$C, were complete after 220 hours. X-ray diffraction (Bruker D8 diffractometer with Cu K$_{\alpha}$ radiation) and neutron diffraction (ILL D20 instrument) indicated that 34\% of the $\mathrm{Cu^{2+}}$ ions in \cual\ are situated on the octahedral sites. The cation distribution of \cuga\ was unable to be determined, however a previous investigation \cite{GON85} found there to be 84\% of the $\mathrm{Cu^{2+}}$ ions situated on the octahedral sites. The chemical structure of both materials is described by the space group ${\it{Fd \bar{3} m}}$, and the lattice parameters are $a$ = 8.263\,\AA\ for \cuga\ and $a$ = 8.045\,\AA\ for \cual.

Magnetic measurements were performed with both dc and ac Quantum Design MPMS SQUID magnetometers. The temperature was measured from 1.8\,K to 300\,K and fields of up to 1\,T were applied. In ac-SQUID measurements the frequency was varied from 50\,Hz to 10,000\,Hz. Specific heat measurements were performed with a Quantum Design PPMS at Pennsylvania State University, USA. Measurements were carried out from 0.5\,K to 30\,K in applied magnetic fields of 0\,T and 1\,T, and corrections were made for the lattice contribution to the specific heat.

\section{Results}
\cuga\ and \cual\ are both found to be spin glasses (in agreement with a previous investigation for \cuga\ \cite{PET01}), and the magnetic properties are very similar for the two materials. It is believed that in the Cu$^{2+}$ based spinels the spin glass state is due to the Jahn-Teller distortions of both the tetrahedrons (A sublattice) and octahedrons (B sublattice) surrounding the copper ions \cite{PET01}. The distortions occur randomly along one of the equivalent 3-fold axes, causing there to be random tetragonal anisotropies in the exchange interactions between ions in the A and B sublattices. Frustration is also present as antiferromagnetic spins are found on the B sublattice, which consists of connected tetrahedra.

Dc-susceptibility measurements as a function of temperature show a divergence of the zero field cooled (ZFC) and field cooled (FC) curves. The susceptibility curves follow the Curie-Weiss law at high temperatures, but the ZFC and FC curves bifurcate at $T \approx $ 2.5\,K for \cuga\ and $T \approx$ 2.2\,K for \cual, due to thermal hysteresis. Magnetisation measurements as a function of field show that as the field is increased the magnetisation does not saturate, and that at low field strengths there is a small hysteresis loop, due to the history dependent nature of the system. These observations are typical for a spin glass state. The spin glass freezing temperature, accurately determined from the derivative of the magnetic susceptibility curve with respect to temperature, is $T_{\mathrm{f}} = 2.89 \pm 0.05\,\mathrm{K}$ for \cuga\ and $T_{\mathrm{f}} = 2.30 \pm 0.05\,\mathrm{K}$ for \cual.

Ac-susceptibility measurements also show characteristic spin glass features. The cusp in the susceptibility that is observed around $T_{\mathrm{f}}$ in zero applied field is suppressed by fields $H \geq$ 0.5\,T. As the susceptibility is measured in increasing ac frequency the spin glass transition temperature, $T_{\mathrm{f}}$, is shifted to higher temperatures. 

\begin{figure}[h]
\includegraphics[width=22pc]{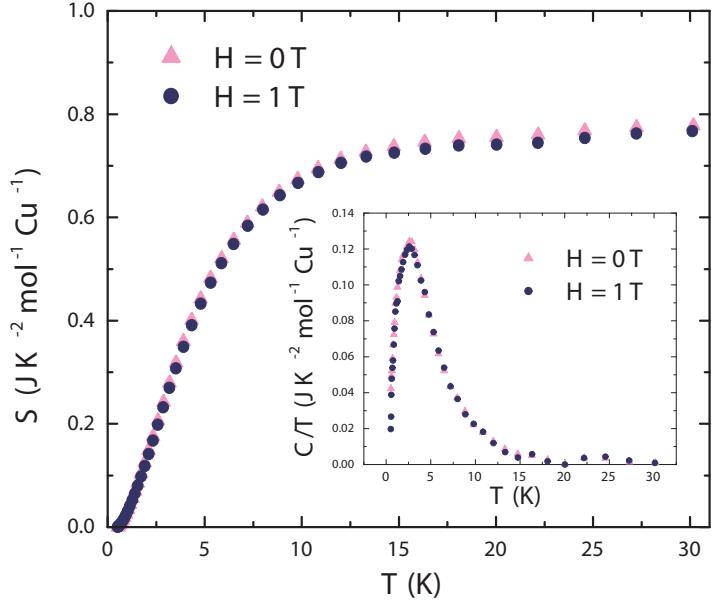}\hspace{2pc}%
\begin{minipage}[b]{11.5pc}\caption{The integrated magnetic entropy of the spin glass \cuga\ from 0.5\,K to 30\,K, in fields of 0 and 1\,T. The inset shows the specific heat divided by temperature of \cuga\ in fields of 0 and 1\,T.}
\end{minipage}
\end{figure}

The magnetic specific heat curves of \cuga\ (shown in the inset of Figure 1) and \cual, measured in fields of 0 and 1\,T, show broad humps around the freezing temperature, indicative of a short-ranged magnetically ordered state. The magnetic specific heat was integrated  from 0.5\,K, in the spin glass regime, to 30\,K, well above the freezing transition at $T_{\mathrm{f}}$, to give the magnetic entropy change, $ \Delta S = \int \frac{C}{T}\,dT $. The expected value of the integrated magnetic entropy for a system with zero entropy at 0\,K is $R\ln2 = 5.76\,\mathrm{JK^{-1}mol^{-1}}$, therefore any difference between this value and the integrated magnetic entropy of the material corresponds to a measure of the zero-point entropy. It is found that for both spin glasses the integrated entropy, shown in Figure 1 for \cuga, is less than $R\ln2$, with values of $1.0\,\mathrm{JK^{-1}mol^{-1}}$ for \cual\ and $0.8\,\mathrm{JK^{-1}mol^{-1}}$ for \cuga. Therefore the values of the magnetic zero-point entropy are $S_{0} = 4.76\,\mathrm{JK^{-1}mol^{-1}} = k_{\mathrm{B}} \times 0.57249\,N$ for \cual\ and $S_{0} = 4.96\,\mathrm{JK^{-1}mol^{-1}} = k_{\mathrm{B}} \times 0.59655\,N$ for \cuga. The ac-susceptibility data indicates that the spin glass state is broken by a field of 1\,T, however the application of a 1\,T field  had little effect on the integrated entropy. Applying a stronger field to the system could result in a more noticeable difference in the entropy.

\section{Discussion}
In the spin glass model proposed by Edwards and Anderson (EA model) \cite{EDW75} the spins are arranged on the sites of a regular lattice, and the interactions between spins are randomly chosen from a (usually) Gaussian distribution. The spins are taken to be Ising, $\sigma_{x} = \pm1$, and the exchange couplings are short range, occurring between nearest neighbour sites only. The EA model correctly predicts some of the observed experimental features of real spin glasses, however it also predicts zero residual entropy. Zero residual entropy is applicable if there is only one possible ground state, or many ground states with a zero probability of a transition between them, but not to a real spin glass with a multi-degenerate ground state \cite{EDW80}.

An infinite-ranged version of the EA model was developed by Sherrington and Kirkpatrick (SK model), using a mean field theory approach \cite{SHE75}. In the SK model of a spin glass the coupling interactions take place equally between all spins in the system (not just nearest neighbours). The spins are Ising spins and the coupling interactions are described by a Gaussian distribution. The SK model also correctly predicts some behaviour of spin glasses, but incorrectly predicts a negative value for the residual entropy, $S_{0} = k_{\mathrm{B}} \times -0.15915\,N = \mathrm{-1.322 \, JK^{-1}mol^{-1}}$. This unphysical value of the residual entropy was improved upon by Parisi, using a replica symmetry breaking scheme, to $S_{0} = k_{\mathrm{B}} \times -0.01\,N = \mathrm{-0.0831 \, JK^{-1}mol^{-1}}$ \cite{PAR79}.

A theory to calculate the degeneracy of the spin glass ground state, based on the infinite-range SK model with Ising spins, was developed by Edwards and Tanaka \cite{EDW80}. It considers there to be many possible accessible configurations of the spins which represent local minimum states at $T = 0$, all of which are degenerate in energy. The number of local minimum states, which are separated from each other by single spin-flip energy barriers, is calculated to be $<g_{0}(\{J\})>_{J} \, = \exp(0.19923\,N)$. The residual entropy of an infinite-range Ising SK spin glass is therefore predicted to be $S_{0} = k_{\mathrm{B}}\ln<g_{0}(\{J\})>_{J} \,= k_{\mathrm{B}} \times 0.19923\,N = \mathrm{1.66\, JK^{-1}mol^{-1}}$. Edwards and Tanaka also calculated the residual entropy of a SK, infinite-range, {\it XY} spin glass to be $S_{0} = k_{\mathrm{B}} \times 0.51691 = \mathrm{4.30\, JK^{-1}mol^{-1}}$ \cite{EDW80}.

The various predictions for the value of residual entropy of a spin glass are all made for ideal, model spin glasses. In real systems it is difficult to emulate all of the required characteristics of a particular model. An example of a system which does conform well to the SK model is a dipolar-coupled Ising magnet with long-range coupling interactions, $\mathrm{LiHo_{0.167}Y_{0.833}F_{4}}$ \cite{QUI07}. The residual entropy of this material was found to be $S_{0} = k_{B} \times 0.18\,N = \mathrm{1.496\, JK^{-1}mol^{-1}}$, which agrees most closely with the prediction for an infinite-range SK Ising spin glass \cite{EDW80}. 

The values of the zero-point entropy for the insulating, spinel spin glasses \cuga\ and \cual\ do not agree particularly well with any of the predictions made for ideal, model systems. The predictions are not made for insulating spin glasses, but for the more usual canonical metallic spin glasses, in which a small number of magnetic ions are randomly doped into a non-magnetic host lattice. \cuga\ and \cual\ also differ from the model systems as they have only short ranged exchange interactions of certain values, instead of a Gaussian distribution of $J_{ij}$, and it is not known whether the spins in the spinel spin glasses are of the Ising, {\it XY} or Heisenberg type. Therefore, no models that have been proposed so far are applicable to these spin glasses, however the values of the zero-point entropy are closest to the prediction for a Sherrington-Kirkpatrick infinite-range, {\it XY} spin glass \cite{EDW80}, $S_{0} = k_{\mathrm{B}} \times 0.51691 = \mathrm{4.30\, JK^{-1}mol^{-1}}$. The higher values that have been found for the spinel spin glasses indicate that there is an even greater degeneracy in the ground state than for a Sherrington-Kirkpatrick {\it XY} spin glass.

\section{Conclusion}
\cuga\ and \cual\ are found to be spin glasses with spin freezing temperatures of $T_{\mathrm{f}} = 2.89 \pm 0.05\,\mathrm{K}$ and $T_{\mathrm{f}} = 2.30 \pm 0.05\,\mathrm{K}$ respectively. The specific heat of both materials was measured, and integrated to give the recovered magnetic entropy. The zero-point entropy of the materials was found to be $S_{0} = 4.76\,\mathrm{JK^{-1}mol^{-1}} = k_{\mathrm{B}} \times 0.57249\,N$ for \cual\ and $S_{0} = 4.96\,\mathrm{JK^{-1}mol^{-1}} = k_{\mathrm{B}} \times 0.59655\,N$ for \cuga. These values are higher than all of the predictions for the zero-point entropy of a spin glass. However, they are of the same magnitude as the predictions, and agree most closely with that for an SK infinite-range, {\it XY} spin glass, $S_{0} = k_{\mathrm{B}} \times 0.51691 = \mathrm{4.30\, JK^{-1}mol^{-1}}$.

\end{document}